\documentclass[prb,aps,groupedaddress,amsmath,twocolumn]{revtex4}

\usepackage[dvips]{graphicx}
\usepackage{dcolumn}
\usepackage{bm}

\begin{document}

%%%%%%%%%%%%%%%%%%

%\widetext

%\twocolumn
%[\hsize\textwidth\columnwidth\hsize\csname
%@twocolumnfalse\endcsname
%%%%%%%%%%%%%%%%%%

\title{Wave function optimization in the variational Monte Carlo method} 

\author{Sandro Sorella}
%\email[]{sorella@sissa.it}
\affiliation{International School for Advanced Studies (SISSA) Via Beirut 2,4
  34014 Trieste , Italy and INFM Democritos National Simulation Center,
  Trieste, Italy} 
\date{\today}

\begin{abstract}
An appropriate  iterative scheme
for the minimization of the energy,  
based on   the variational Monte Carlo (VMC) technique,
is  introduced and compared with existing stochastic schemes. 
We test the various methods for the 1D Heisenberg ring and the 2D t-J model
and show that,  with the present scheme,
very accurate and efficient calculations are possible, even for
several  variational parameters.
Indeed, by using   a very efficient statistical evaluation of   
the first and the second energy derivatives,  
it is possible to define a very
rapidly converging iterative scheme that, within VMC, 
is much more convenient than the standard Newton method. 
It is also shown how to optimize simultaneously
both  the Jastrow  and the determinantal  part of the wave function.
\end{abstract}


\maketitle
%]
%%%%%%%%%%%%%%%%%%
%%%%%%%%% PACS USED
%%%%  02.70.Ss  Quantum Monte Carlo Methods
%%%%  71.10.Fd Lattice fermion models (Hubbard model, etc.)
%%%%  31.10.+z Theory of electronic structure, electronic transition and chemical binding
%%%%%%%%%%%%%%%%%%
%\narrowtext
 Since the seminal work of Gutzwiller, 
 a large amount of progress has been achieved in the  understanding 
of strongly correlated materials  by means of correlated wave functions (WF). 
The fractional quantum Hall effect\cite{laughlin} 
 and the theory-still controversial- 
of high temperature superconductors\cite{rvb}
 are two important among many other examples. 
 
Most  correlated WF, here indicated by $|\psi_G\rangle$, 
can be  obtained\cite{jain,gros,giamarchi} 
by applying a two body Jastrow factor $J$ to a  mean-field WF  
$|D\rangle$ described by a  single determinant or a  
 superconducting BCS  state (SDBCS), so that: 
\begin{equation}
|\psi_{G}\rangle  = J |D\rangle 
\end{equation}
where $J$ contains  two-body density-density 
and spin correlations.
%\cite{pippo}
%, as e.g. in the Gutzwiller WF the factor $J$ suppress 
%energetically  too expensive configurations with doubly occupied sites.
For accurate calculations, the inclusion of these  two-body correlations 
in $\psi_G$ is possible by means of 
 the VMC statistical approach. 
%that 

% until now  is certainly much less efficient than  standard mean-field or 

%  Density-Functional theories. 
However  the application of correlated WF to complex electronic 
systems\cite{filippi,filippispread,steepest}, 
 or-e.g.-to describe correctly  the metal-insulator transition\cite{capello}, 
%or to understand the role of impurities
% in correlated superconductors\cite{rice,ogata},
requires  the use of many variational parameters. A drawback 
that has certainly 
limited the application of VMC to important physical and chemical 
problems with a large number of electrons.

In order to overcome the above limitations  of VMC,  in this
 paper we introduce a  minimization scheme, that allows  to optimize 
efficiently the energy expectation value:
\begin{equation} \label{energy} 
E_{\bf \alpha}={  \langle \psi_{\alpha} | H | \psi_{\alpha} \rangle  
\over \langle \psi_{\bf \alpha} | \psi_{\bf \alpha}  \rangle }  
\end{equation}
where $H$ is the model Hamiltonian considered, 
 ${\bf \alpha} = \alpha_1,\alpha_2,\cdots \alpha_p$ is a set 
of $p$  variational parameters appearing in  the WF 
$\psi_{\bf \alpha}(x)=\langle x| \psi_{\bf \alpha} \rangle$, where 
 $|x\rangle $ denotes a configuration with defined  electron spins and 
positions. Henceforth the symbol $< >$ indicates the 
quantum expectation value over the state $\psi_{\bf \alpha}$, 
so that $E_{\bf \alpha}=<H>$. 
 Within the Newton method (NM),  in order to improve   the  
variational  parameters, 
$E_{\bf \alpha+ \gamma}$ is expanded   
quadratically in $\gamma$ 
and minimized, then the parameters are  changed 
${\bf \alpha^\prime}= {\bf \alpha}+{\bf \gamma}$, and the   
iteration repeated   until $\gamma$ is negligible.   
In the following we describe an iterative technique, that is based on  
a similar strategy, and that, within VMC, performs 
much better than standard  NM. 

At each iteration and for each variational parameter $\alpha_k$  
we define\cite{casula} an  operator $O_k$, diagonal in the 
configuration basis,  with diagonal elements: 
\begin{equation}  \label{operators}
O_k(x)=  
{\partial_{\alpha_k} \psi_{\bf \alpha} (x)  \over \psi_{\bf \alpha} (x)}. 
\end{equation}
%%%%%%%%%%%%%%%%%%%%%%%
%Notice that  all these  operators, at the minimum energy for 
%$\psi_{\bf \alpha}$,  
% satisfy the same stationary condition 
%\begin{equation} \label{stationary}
%\langle  O H \rangle - \langle O \rangle \langle H \rangle = 0 
%\end{equation}
%valid for any operator in the exact 
%ground state (GS)  of $H$.
%Thus at the minimum energy the expectation value of these operators are 
%at least consistent with an exact property of the GS, and therefore 
%more sensible results for correlation functions $<O_k>$  are expected. 
%%%%%%%%%%%%%%%%%%%%%%%%%%%%%%%%%%%%%%%%%%%%%%%%%%%%%%%%%%%%%%%%%%%%%%%%
%% We emphasize here that, the reason to include the Jastrow term 
%$J = exp[$Density-Density correlations (+/ or  Spin-Spin correlations )$]$      in a many-body WF, is not only meant to gain a bit of energy, but 
%%more importantly to correct and sometimes to  describe qualitative
%% new effects, beyond the Hartree-Fock  paradigm, especially when 
%these correlations cannot be,  not even qualitatively, described by this 
%theory. 
%%%%%%%%%%%%%%%%
In order to simplify the derivation 
we assume that  these  operators $O_k$ appear  
in  the WF in the simple exponential form 
${\rm exp} [ \sum_k \alpha_k (O_k-<O_k>)]$.
This is certainly the case 
for  the Jastrow  factor $J$, where 
 the operators $O_k$ are just 
 density-density or spin-spin correlations.
 Here 
for convenience we have  subtracted from each $O_k$ 
the corresponding  average value $<O_k>$, which provides  
an irrelevant multiplicative  constant in the WF. 
For the variational parameters 
corresponding to   the uncorrelated part $|D\rangle$,   
this exponential form is not exactly fulfilled,  
but  is valid only for small changes of the variational parameters 
$|\psi_{\bf \alpha+\gamma}\rangle  \propto  {\rm exp} [ \sum_k \gamma_k (O_k-<O_k>)] 
|\psi_{\bf \alpha}\rangle,$
namely within linear order in $\gamma$. 
In such a case, this expression
does not provide the exact second
derivative of the WF with respect to $\gamma$. Although this information
is in principle required for the NM, we will show that
the accurate determination of these
terms  is not really important within VMC,  because  the standard  NM
is rather inefficient even when these  second WF derivatives
are computed exactly.
To this purpose we write
$|\psi_{\bf \alpha+\gamma}\rangle$ in a slightly more general form that 
depends on a further parameter $\beta$:
\begin{eqnarray} \label{linear}
|\psi_{\bf \alpha+\gamma } \rangle& \simeq& 
[ 1 + \sum\limits_{k} \gamma_k (O_k-<O_k>)   \\
+  &{\beta\over 2}&  \sum\limits_{k,k^\prime} 
 \gamma_k \gamma_{k^\prime} (O_k-<O_k>)
(O_{k^\prime}-<O_{k^\prime}> ) ]  | \psi_{\bf \alpha} \rangle \nonumber  
\end{eqnarray}
 As discussed before, this expansion 
 is valid only within linear (quadratic) 
order in $\gamma$ for $\beta \ne 1$ ($\beta=1$ Jastrow case).
We consider this more general form because the value of $\beta$ can 
be used to improve the efficiency of the minimization scheme within the VMC. 
On the other hand it is clear that,  as the minimum is approached,   
the non linear contribution proportional to $\beta$, as well as the 
higher order ones,  become 
negligible and irrelevant for the WF optimization.

By substituting expression (\ref{linear}) in Eq.(\ref{energy}), 
the energy can be systematically 
expanded in powers of $\gamma_k$:
\begin{equation} \label{form}
\Delta E= - \sum_k \gamma_k f_k + 
{1\over 2} \sum_{k,k^\prime} \gamma_k \gamma_{k^\prime} 
 \left[S_h+(1+ \beta) G\right]^{k,k^\prime} 
\end{equation} 
where:
\begin{eqnarray}
 S_h^{k,k^\prime} &=&  < 
 \left[ O_k, \left[ H, O_{k^\prime} \right] \right] >   
 \label{defshmat}  \\
G^{k,k^\prime} &=&  2 < (H-E_{\bf \alpha}) (O_k-<O_k >) 
(O_{k^\prime}- < O_{k^\prime}> ) >   \nonumber \\ 
%     &+&    <  (O_k-<O_k >)
%(O_{k^\prime}- < O_{k^\prime}> ) (H-E_{\bf \alpha}) > \nonumber  \\
f_k &=& -\partial_{\alpha_k} 
  E_{\bf \alpha}=   -2  < (H-E_{\bf \alpha} )  O_k > 
\label{quantumdef}
\end{eqnarray}
In the above equations we have used the 
hermitian character  of all the operators involved, 
implying  for instance  that  $<O_k H >=<H O_k>$;   
  $f_k$ indicate  the  forces
acting on the variational parameters and  vanishing  
at the minimum energy condition,  
  $S_h^{k,k^\prime}$  
represent the excitation matrix elements corresponding to the 
operators $O_k$,   
$G^{k,k^\prime}$ take into account the remaining contributions 
appearing when the WF is not exact ($<H-E_{\bf \alpha}>  \ne 0$),
whereas the square brackets  indicate the commutator.
The WF parameters  can be  then 
iteratively changed  $\alpha_k \to \alpha_k + 
\gamma_k$, by minimizing Eq.(\ref{form}) 
whenever 
$S_h+(1+\beta) G$ is positive definite. 
%(has all positive eigenvalues,  
%a condition that generally  holds  close enough to the minimum)
The minimum energy is obtained for:  
\begin{eqnarray}
{ \bf \gamma } &=&  B^{-1}  {\bf f}  \label{iter}  ~~~ {\rm with:}\\
B &=&  S_h+(1+\beta) G  \label{defb}
\end{eqnarray}   

If  $B$ is not positive definite, 
 the quadratic form (\ref{form}) is not bounded from below, meaning that  
 higher order terms   are important in the expansion (\ref{form}). 
In this case  the correction (\ref{iter}) may lead to a  higher  energy rather 
than to a lower  energy WF.
In order to overcome this difficulty, 
we change   the matrix $B \to B + \mu S$
where $S$ is  the  positive definite\cite{notesr}  overlap matrix:
\begin{equation} \label{defs}
S^{k,k^\prime} =< (O_k-<O_k>) (O_{k^\prime} - <O_{k^\prime}> ) >
\end{equation} 
Similarly to  the previous 
 Stochastic Reconfiguration  technique\cite{casula}, 
we use the same matrix $S$   
in order to have a well defined minimum 
for the energy $\Delta E$ in all cases. This is achieved 
by imposing the constraint that the  linear 
 WF change  $\Delta {\rm WF}= 
 (|\psi_{\bf \alpha+\gamma} \rangle -|\psi_{\bf \alpha }  \rangle)/  | \psi_{\bf \alpha} |$,  obtained with $\beta=0$ in Eq.\ref{linear}, 
  cannot be larger than a certain amount $r$.
  Hence  the control parameter $r$ is  defined by means of the inequality: 
\begin{equation} \label{control}
|\Delta {\rm WF}|^2 \le  r^2
\end{equation} 
where, from   Eqs.(\ref{linear},\ref{defs}), 
 $|\Delta {\rm WF}|^2= \sum_{k,k^\prime}  \gamma_k \gamma_{k^\prime} S^{k,k^\prime}$. 
This constraint 
allows to work always with a positive definite matrix 
$B$ and,  for small enough $r$, 
the energy is {\em certainly} lowered by changing the parameters 
according to (\ref{iter}).  
The constant  $\mu\ge 0$  is non zero 
whenever  (\ref{defb})  is non positive definite or 
$|\Delta {\rm WF}|$   corresponding to (\ref{iter}) 
exceeds $r$. In these cases the constant $\mu$  
 is obtained  as  a Lagrange 
multiplier, namely by  minimizing 
 $\Delta E + \mu |\Delta {\rm WF} |^2$,  
with the condition    
$|\Delta {\rm WF}|=r$, which 
is simple to fulfill with standard iterative schemes. 
A similar control parameter was used in Ref.\onlinecite{filippi},
by adding the identity matrix $I$  scaled by $a_{diag}>0$ to the Hessian,  
namely $B=S_h+ 2 G + a_{diag} I$, so that for $a_{diag}\to \infty$ 
the steepest descent is obtained, whereas 
within our scheme, for   $\mu \to \infty$ 
the stochastic reconfiguration (SR)   technique is recovered 
with $B=S/\Delta t $ and $\Delta t=1/\mu$ small.\cite{casula}

As it is clear from its definition  (\ref{quantumdef}) the matrix 
$G$ is zero when $\langle \psi_{\bf \alpha}|$ 
coincides with the exact ground state,  because in this case  
$\langle \psi_{\bf \alpha} | (H-E_{\bf \alpha})=0$.
Therefore this matrix 
should be very small compared with  $S_h$   
for a good variational WF.  
This implies that, by changing   $\beta$ in Eq.\ref{defb}, 
we can obtain iterative  methods 
converging to the minimum energy  with approximately the same {\em  small} 
number of iterations. 
 As we will see in the following, for the optimization of the Jastrow (Jastrow 
and SDBCS)   it is particularly convenient to use $\beta=-1$ ($\beta=0$), 
so that $B=S_h+\mu S$ ($B=S_h+G+\mu S$), 
with $\mu$ determined by (\ref{control}) when different from zero. 
Henceforth the $\beta=1$ technique  is  indicated by 
 NM, namely  the standard Newton method used 
in Refs.(\onlinecite{filippi,rappe}),  whereas the one 
defined by an appropriate choice 
of the  parameter $\beta \ne 1$, 
is named  SR with Hessian 
acceleration (SRH).

\noindent{\em  Statistical averages in VMC.}
 In VMC the statistical averages  
required for each iteration are evaluated 
over $M$  sampled configurations $\{ x_i \}$-named bin-, obtained after 
a small equilibration, and  distributed according to the square of the WF.    
We indicate the corresponding statistical averages with  
the symbol  $ <<  >>$, e.g. 
$ << O_k (x) >> = 1/M \sum_i O_k (x_i)$, so that 
for large $M$ $<< O_k (x) >>$ 
 coincides with the exact quantum average $<O_k>$, 
with a statistical accuracy $\propto {1\over \sqrt{M}}$.

After each bin we change  the variational 
parameters according to Eq.(\ref{iter}), by using the appropriate matrix $B$ 
discussed before. 
The SRH is stable   for small enough   $r$ 
and  may  converge much faster than  SR.
With a  larger  bin length $M$, the   value of $r$ can  be substantially increased, leading therefore to a much faster convergence within the SRH 
minimization scheme.  

As also noted in Ref.\onlinecite{filippi} 
 it is extremely  important that the quantities 
evaluated within the statistical  approach are 
obtained by means of  fluctuations $\delta A(x)= A(x) - << A(x) >> $ 
of suitable variables $A(x)$, that depend only  
on the  electronic configuration $x$ sampled with  VMC.
Therefore, it is useful to  express 
the forces $f_k$,  the matrices $S_h$ and  
 $S$  by appropriate fluctuation averages: 
\begin{eqnarray} \label{allcorrs}
f_k &=& -2  <<\delta  e_L(x) \delta   O_k(x) >>  \label{defforza}  \\
 S_h^{k,k^\prime} &=&   << \delta \partial_{\alpha_k} e_L (x) \delta 
 O_{k^\prime}(x)>>
+ (k\leftrightarrow k^\prime)     \label{defsh} \\
S^{k,k^\prime} &=& << \delta O_k (x) \delta O_{k^\prime}(x)>>  \label{defss} \\
G^{k,k^\prime} &=& 2  
 << \delta e_L(x) \delta O_k(x) \delta O_{k^\prime} (x)  >>
\label{formg}
\end{eqnarray} 
where 
  $e_L(x) = { \langle \psi_{\bf \alpha} | H | x \rangle \over 
\langle \psi_{\bf \alpha} | x \rangle }$ is the so called 
local energy. In order to derive Eq.(\ref{defsh}) 
 we insert a  completeness
 $\sum_x | x \rangle \langle x | = I$ in the RHS of 
Eqs.(\ref{defshmat}), and we  
 also make use of the identity
$\langle x  |H O_k |\psi_{\bf \alpha} \rangle = 
( \partial_{\alpha_k}  e_L (x) + O_k(x) e_L(x) ) \psi_{\bf \alpha} (x) $. Then 
we notice  that   $<<\partial_{\alpha_k} e_L (x) >>=0$
 for $M\to \infty$,\cite{caffarell}
so that the fluctuations of $\partial_{\alpha_k} e_L(x)$ 
can conveniently appear in Eq.(\ref{defsh}).
 This way  to  evaluate $S_h$ 
is affected by much smaller statistical errors   than  
the previous estimate\cite{rappe} 
$<<  \partial_{\alpha_k} e_L (x) 
 O_{k^\prime}(x)  >> +   (k\leftrightarrow k^\prime).$
% requiring    
%a  much larger  $M$, for the same statistical accuracy. 
For the same reason 
 the matrix $B$,  used in the iteration (\ref{iter}),  is  affected by 
the large noise present in $G$. 
Indeed, for $\beta =-1$, $B$ does not depend  on $G$ and the corresponding   
statistical fluctuations  are much reduced, especially 
for a large number of electrons $N_e$. 
In fact the operators $O_k$ and the local energy scale with $N_e$, 
their  fluctuations 
$\delta e_L, \delta O_k$ 
  in (\ref{formg}) with  $\sqrt{N_e}$, whereas the corresponding 
statistical averages for $G$ are  of order $N_e$, because  
they contribute  to an extensive energy in (\ref{form}). 
This implies  {\em zero} 
  signal to noise ratio ($ \simeq { 1\over  \sqrt{N_e} } $) 
  in the matrix $G$ (or similarly in $S_h$ without using 
the correlation \ref{defsh}) for large $N_e$  
 compromising the  efficiency of the  VMC.
On the other hand 
it is clear that both the matrices  $S$ and $S_h$  
in Eqs.(\ref{defsh}-\ref{defss}) 
 are  much better behaved, because they are  
obtained by averaging statistically quantities of order $N_e$, that  
 have a mean value  of the same order.

Unfortunately, for the optimization of $|D\rangle$, 
uncontrolled divergences in the evaluation of $S_h^{k,k^\prime}$
appear when both indexes $k$ and $k^\prime$ correspond to 
the SDBCS parameters.
This is  due to 
the zeros of  $\langle x|D\rangle$, 
implying that   
 $\partial_{\alpha_k}  e_L(x)\simeq - O_k(x) e_L(x)$ 
and $O_{k^\prime} (x)$ 
wildly fluctuate in Eq.(\ref{defsh})   
because they can both diverge for $\psi_{\bf \alpha} (x) \to 0$. 
% just when $k$ and $k^\prime$ correspond to the 
% SDBCS matrix elements. 
The way to overcome this difficulty is to use $\beta=0$ in Eq.(\ref{defb}) 
since the most 
relevant divergences in $S_h$ are exactly canceled by $G$ {\em just 
for this value of $\beta$}.
In principle the matrix $B$ {\em cannot} be evaluated statistically
for $\beta\ne 0$
 because the statistical uncertainty remains even 
for $M\to \infty$, due to these  large fluctuations. 
 For  $\beta=0$ and  for large number of electrons $N_e$, 
the problem of 
zero signal to noise ratio in $B$ cannot be avoided
for Jastrow and SDBCS  optimization,
but can be substantially alleviated  by a good WF  choice, because 
$G$ has the zero variance property:  $G=0$ without noise, 
whenever $\psi_{\bf \alpha}$ is an exact eigenstate.

\noindent {\em Results}
We have tested the simple SR method, the presently discussed 
 SRH one, and the 
recently introduced  NM\cite{filippi}  
($\beta=1$ in our notations) 
on  simple $L-$sites lattice models,  the 1D Heisenberg model (1DHM) 
 and the t-J model\cite{superb}, 
 because  we believe they  represent  useful 
benchmarks  to compare various techniques.
The Jastrow factor of the initial WF was set to zero 
   for both models.
Moreover, consecutive configurations  in 
 each  bin are separated by   
$2\times L$ Metropolis attempts.  

For the 1DHM we use   a   WF  
containing  a long range Jastrow factor in the form: 
$ exp (1/2  \sum_{i,j} v^z_{i,j} S^z_i S^z_j ) |D>$  where 
$|D\rangle$ is a N\'eel state  with 
magnetization along the $x-$spin axis.
%,  or a $BCS$ state 
%($|BCS\rangle$) defined by a gap  function 
%$\Delta_k = \Delta_1 cos(k) + \Delta_2 \cos (3 k)$ containing two 
%variational parameters.
Here $\vec S$ are usual  spin-$1/2$ operators and projection 
over  zero total spin$-z$ component   
is   also assumed.
The long range Jastrow is essential 
to  destroy the long range magnetic 
order in 1D and obtain a sensible and accurate ansatz\cite{manousakis}.
By using all spatial symmetries of the model, 
we remain with  $L/2-1$ independent parameters in the Jastrow factor.

\begin{figure}
\includegraphics[width=\columnwidth]{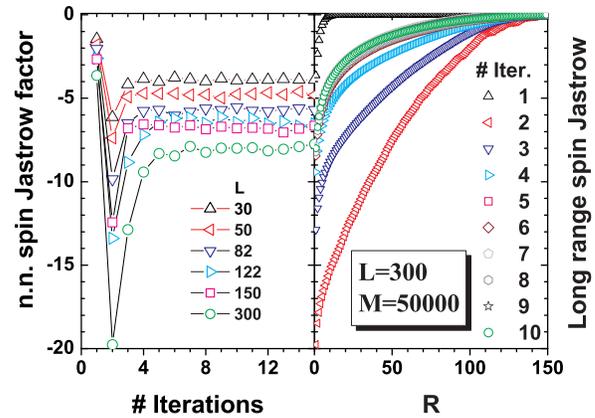}
\vspace{-0.5cm}
\caption{\label{alljasvsiter}
Left: convergence of the SRH technique for different sizes.
Right: the same for all the $149$ independent VMC parameters 
for the maximum size studied. 
For all simulations  the bin length (the control parameter $r$ 
defined in Eq.(\ref{control}))  $M$ was  taken proportional to $L^2$ ($L$), 
starting from $M=500$ ($r=2$) for $L=30$. 
}
\end{figure}

As shown in Fig.(\ref{alljasvsiter}), with the present SRH technique,  
a  converged value  of the nearest neighbor $v^z$  
is obtained  in  less than  $10$ iterations for 
all sizes studied (similar convergence is obtained at all distances, 
see right panel),
 implying that the use of the matrix $S_h$ ($\beta=-1$ 
has been  used in this case) 
is  very effective to accelerate the convergence to 
the minimum energy WF. 
By the SR technique, stability forces a very slow convergence:
%For the previous technique, the simple SR,
%the value of $J \Delta t$, cannot be taken 
%larger than  $0.125$ for a stable but slow convergence. 
% The SR  scheme eventually converges  approximately to the 
% same value after 
%$200$ iterations (not shown).
% The larger is $\Delta t$, the  faster is the SR method, but 
%$\Delta t$ cannot be too large.
%  Though the maximum possible 
%$\Delta t$ appears almost independent of $L$,
% Unfortunately, within SR,  
 the required number of 
iterations increases with  
the system size $\propto  L^2$  because the model  is gapless   
and accurate  calculations become prohibitive already for 
$L\simeq 100$.
%%%%%%%%%%%%%%%%%%%%%%%%%%%%%%%%%%%%%%%%%

%%%%%%%%%%%%%%%%%%%%%%%%%%%%%%%%%%%%%%%%%
\begin{figure}
\includegraphics[width=\columnwidth]{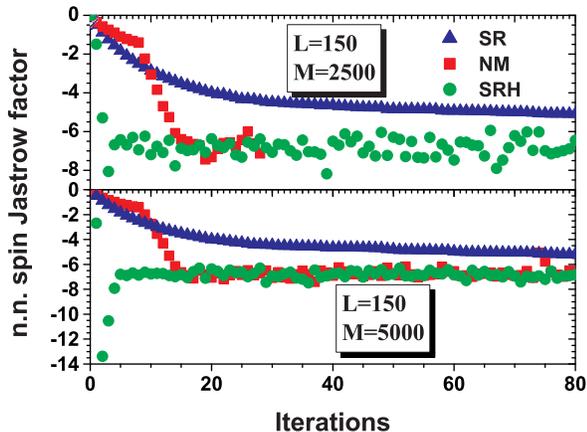}
\vspace{-0.5cm}
\caption{\label{compare}
Convergence of the VMC n.n. spin Jastrow for various methods (see text).
For the SR technique  $\Delta t J=0.125$. 
For the NM  the first 4 iterations are with $a_{diag}=0.02$, 
for the next 6 $a_{diag}=0.002$ and the rest with $a_{diag}=0$.
For $M=2500$ the NM   is  unstable after $~30$ iterations 
due to a big fluctuation of the matrix $G$ (see text) not used  in SRH.
For the SRH technique $r=5$ for $M=2500$ and $r=10$ for $M=5000$.
After the third iteration  $|\Delta {\rm WF}| < r $ is always satisfied, even after 
further $1000$ iterations (not shown). 
}
\end{figure}
As shown in Fig.(\ref{compare}) 
SRH is  remarkably faster than  SR. 
After convergence, with much longer runs,  
SR and NM  appear  $\gtrsim 50$ and $\gtrsim 10$ 
 times less efficient than SRH respectively,  for 
obtaining  the variational parameter within a given statistical error, 
by performing a statistical average,  
as  discussed in Ref.\onlinecite{casula}. 
SR  has  a much larger correlation time  and NM is affected by 
 larger fluctuations. 
%Moreover  for the SR the time spent for equilibration may become a relevant 
%fraction  of the calculation, and the so large correlation time $N_0$ prevent 
%to estimate accurately the corresponding statistical error.
%For the case shown in the picture,  by continuing  
%all  the simulations  with  further $N=1000$ iterations and $M=5000$,  
% the standard deviations  $\sigma $ ($\propto {1  \over \sqrt{N/N_0}} $) 
% of the variational parameter   
%for  SR, NM , and SRH, are 
%($0.16(2)$, $0.12(2)$ and $0.069(2)$)
%$0.05(1)$,  $0.005(1)$, $0.0018(2)$ respectively.
% Thus both 
% SRH and NM clearly improve over  
% the simple SR, e.g. in this case,  for  a given value of $\sigma$,   one needs 
%$\simeq 700$ times and $\simeq 100$ times smaller $N$, respectively, which is 
%remarkable  because the extra computational cost for computing $S_h$ is almost 
%negligible (for $N_e=30$ all methods are equally efficient).   
The performances of SRH  appear always optimal just for $\beta=-1$, 
especially for large 
$N_e$,   due to the argument presented below Eq.(\ref{formg}).
%It is interesting that 
%even for an high quality WF such as the one studied. this increased   
%efficiency of SRH compared with NM,   can be already resolved  for $L=150$, 
%because the rms of the variational parameters are about a factor two smaller in this case. 
This value of $\beta$   allows  also  to use 
small  bin length  (see Fig.\ref{compare}) and to be very  efficient  
far away from convergence.

\begin{figure}
\includegraphics[width=\columnwidth]{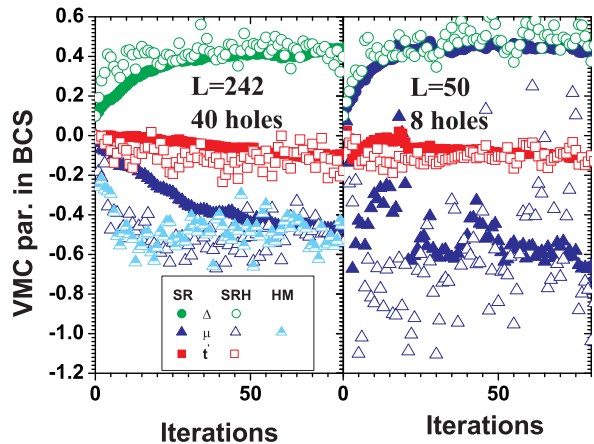}
\vspace{-0.5cm}
\caption{\label{comparedet}
 Parameters in $|D\rangle$, while optimizing the Jastrow part simultaneously.
for the $t-J$ model at  $J/t=0.4$ with bin $M=5000$.
 For SRH the control parameter is  $r=0.5$ ($r=0.1$) for $L=242$ ($L=50$), 
for SR $J \Delta t=0.125$. The hybrid method (HM)
 is also shown (left) for $\mu_0$  ($r=1$ $J \Delta t =1.25$).
}
\end{figure}

\noindent {\em Optimization of  $|D\rangle $}. We have also tested the SRH technique 
in  the  2D $t-J$ model at $J/t=0.4$\cite{superb} 
for optimizing 
the parameters contained in the $D=|BCS\rangle$
 where the BCS Hamiltonian defining the $BCS$ 
state  has a gap function 
$\Delta_k=\Delta (\cos k_x -\cos k_y)$ and a dispersion band     
$\epsilon_k = -2  ( \cos k_x + \cos k_y) -4 t' \cos k_x \cos k_y -\mu_0$, 
where $\mu_0,t'$ and $\Delta$ are the three variational parameters,  
that are optimized together with all the forty (ten) independent 
ones contained in the density-
density Jastrow factor $exp ( 1/2 \sum_{i,j} v_{i,j} n_i n_j )$ 
for $L=242$ ($L=50$). 
In this case, as discussed before,  
$\beta=0$ has to be used, namely $B=S_h+G$ 
in the iteration scheme (\ref{iter}). 
Moreover, in contrast to  the Jastrow case, the condition (\ref{control}) cannot
 be always verified with  $\mu=0$  after equilibration, due to 
occasionally large  fluctuations in the  matrix $B$. 
Nevertheless,  the SRH remains stable, by using  a   small enough value of 
 $r$. In this way,   
 these  large fluctuations of $B$ are very  efficiently damped by condition 
(\ref{control}). 
%In order to achieve a more stable convergence, 
%  it is  also useful 
%to multiply the parameter corrections ${\bf \gamma}$ 
%in the iteration (\ref{iter}) 
%by a constant $\eta < 1$, so that (\ref{control}) can be more easily fulfilled 
%after equilibration at an expence of a larger correlation time 
%$N_0 \to N_0/\eta$.
% Henceforth we use $\eta=0.25$.
 As shown  in Fig.(\ref{comparedet}), also 
in this case 
 the SRH  provides a substantial reduction of 
the iterations required for convergence. 
 As expected  this is accompanied also by a sizable 
increase of the fluctuations in the parameters especially for the chemical 
potential $\mu_0$.
% Similar to the previous case, by continuing the simulations for 
%further 1000 iterations, it turns out  that the SR and SRH 
%are  equally efficient for determining 
% the variational parameters for $L=50$, the former having smaller fluctuations 
%and the latter a shorter correlation time $N_0$. 
Indeed,  though the efficiency
 increases   with $L$, in favor of  SRH.
% with similar 
%(or even smaller for $\mu$) 
%fluctuations for the parameters. 
%Thus  for $L\ge 242$ 
%the SRH allows to reduce, by a considerable  fraction,  the 
%correlation time $N_0$,
this improvement  is not  
as  important as in the previous case,  
 for the optimization of the Jastrow part alone. 

For  the simultaneous optimization of the Jastrow and SDBCS 
parts of the WF, 
  a more efficient scheme  is 
to use  the SR 
for  the SDBCS parameters  and the SRH with $\beta=-1$ for the remaining ones 
 (hybrid method in Fig.\ref{comparedet}).
In this way, in the iteration (\ref{iter}), 
  the noisy and computationally expensive ($\simeq 5$ times more)  
SDBCS matrix elements  $S_h^{k,k^\prime}$  
are replaced by  $S^{k,k^\prime}/\Delta t$, whereas all the other matrix elements 
of $S_h$ are evaluated including also  
the ones  coupling  $J$ and $|D\rangle$. For these matrix elements  
an   estimator  different from the symmetric one (\ref{defsh}) 
 is used for $S_h$: 
$S_h^{k,k^\prime}=S_h^{k^\prime,k}=
 2  << \delta \partial_{\alpha_k} e_L (x) \delta
 O_{k^\prime}(x)>>$, involving only  local energy derivatives of 
Jastrow parameters. 
Afterall   suitable values  of $r$ and $\Delta t$ 
provide performances similar to  SRH for the size studied, 
with a much cheaper computational 
cost per iteration and  with a signal to noise ratio 
 certainly stable for large $N_e$, as only $S^h$ and $S$ (and not $G$)  
are  used in this case. 
%%%%%%%%%%%%%%%%%%%%%%%%%%%%%%%%%%%%%%%%%
%%%%%%%%%%%%%%%%%%%%%%%%%%%%%%%%%%%%%%%%%

% The use of the hessian matrix in VMC  
% should allow a dramatic reduction of the number of iterations, 
%probably by several orders of magnitude 
%for obtaining   
%converged results in a single particle  basis expansion of a  
%realistic WF\cite{filippispread}.  
%In fact  the efficiency of SR or steepest descent, 
% is limited by the condition that  $\Delta t ^{-1}$ 
%has to be  smaller than the high energy cut-off, 
%that is obviously diverging when this  basis approaches 
% completeness\cite{filippispread}. 
%whereas this cut-off is bounded by 
%$\simeq 4 J$ ($\simeq 20 J$)  -independent of $L$- 
%for the Heisenberg  ($t-J$)   model considered.    

In conclusion, the  optimization methods  we have described, based on a very efficient VMC evaluation of the Hessian matrix, open the possibility to study 
electronic systems with correlated WF containing many variational 
parameters,  and  with  an efficiency that is now comparable with 
non statistical methods of some time ago\cite{cp}.

%because, within the 

%present iterative methods, the time required for optimizing 

%many parameters, does not cost more than a single simulation at fixed 

%variational parameters, thus compensating for a big factor, the number 

%of iterations, necessary for deterministic methods to converge. 

 I acknowledge useful comments by  
 A. Parola.  M. Casula, F. Becca, C. Filippi, and  C. Umrigar. 
  This work was partially supported by COFIN MIUR-2003.

%Unused bibitems

%\bibitem{noder} We neglect in this case the second derivatives of the WF 
%with respect to the variational parameters.
%Unused bibitems

%
%
%
%
%Unused bibitems

%\bibitem{sorella} S. Sorella \prb {\bf 64} 024512 (2001).


\begin{thebibliography}{99}
\bibitem{laughlin} R. B. Laughlin, \prl {\bf 50}, 1395 (1983).

\bibitem{rvb} P. W. Anderson, Science {\bf 235}, 1196 (1987).

\bibitem{jain} G. Dev, J. K. Jain \prb {\bf 45}, 1223 (1992).

\bibitem{gros} C. Gros \prb {\bf 38}, R931 (1988).

\bibitem{giamarchi} T. Giamarchi, C. Lhuillier \prb {\bf 43}, 12943 (1991).

\bibitem{filippi} C. J. Umrigar,  C. Filippi to appear in \prl. 

\bibitem{filippispread} F. Schautz, C. Filippi J. Chem. Phys. 
{\bf 120}, 10931 (2004). 






\bibitem{steepest} A. Harju {\it et al.}, Phys. Rev. Lett. \textbf{79}, 1173 (1997).


\bibitem{capello} M. Capello {\rm et al.}, \prl {\bf 94}, 026406 (2005).

\bibitem{casula}  M. Casula, C. Attaccalite, S. Sorella 
J. Chem. Phys. {\bf 121}, 7110  (2004). 

\bibitem{notesr} When the lowest eigenvalue of 
$S$ is very small, for a stable method, the  
irrelevant parameters of the WF have to be  removed,  e.g.  with the method 
described in Ref.\onlinecite{casula}. 

\bibitem{rappe}  M. Casalegno, M. Mella,  A. M. Rappe,
  J. Chem. Phys. \textbf{118}, 7193 (2003).

\bibitem{caffarell} R. Assaraf, M. Caffarel \prl {\bf 83}, 4682 (1999).

\bibitem{superb} S. Sorella et al. \prl {\bf 88}, 117002 (2002).

\bibitem{manousakis}  see e.g. E. Manousakis, \rmp  {\bf 63}, 1 (1991).

\bibitem{cp} R. Car and M. Parrinello \prl {\bf 55}, 2471 (1985).

%\bibitem{iteration} To this purpose the generalized diagonalization of 

%the matrix $S^h$ is used: 

%$ \sum_{k^\prime}  S_h^{k,k^\prime} \psi^i_{k^\prime} 

% = \lambda_i \sum_{k^\prime} 

%S^{k,k^\prime}  \psi^i_{k^\prime} $, where $\lambda_i$ 

%and $\psi^i_k$ are generalized 

%eigenvalues and corresponding eigenvectors satisfying $\sum_{k,k^\prime} 

%S^{k,k^\prime} \psi^i_k \psi_{k^\prime}^j=\delta_{i,j}$.

%Thus

%%$|\Delta WF|^2 =\sum_i [v_i /(\lambda_i + \mu)]^2$  can be easily computed 

%for several  values of $\mu$,  once   $v_i=\sum_k \psi^i_k f_k$ is evaluated 

%at the beginning, 

%\bibitem{steepest} A. Harju, B. Barbiellini, S. Siljamaki, R.M. Nieminen and

%G. Ortiz, Phys. Rev. Lett. \textbf{79}, 1173 (1997)

%\bibitem{vmcuberalles} M. Snajdr and S. M. Rothstein, 

%J. Chem. Phys. {\bf 112}, 4935 (2000).

%\bibitem{notevariance} Strictly speaking also $f_k$ and $S$ in (\ref{formg}) 

%have infinite variance, nevertheless  self averaging applies, and 

%the statistical averages converge for $M\to \infty$.

% The 

%Stochastic approach is therefore well defined in this limit.  

\end{thebibliography}
\end{document}